\documentclass[usenatbib,usegraphicx,useAMS]{mn2e}
\usepackage{mnoverride}
\usepackage{setspace}
\usepackage{soul}
\usepackage{deluxetable}
\usepackage{amsmath}                
\usepackage{amsfonts}               
\usepackage{amssymb}                
\usepackage{multirow}
\usepackage[usenames,dvipsnames]{color}
\usepackage{mathptmx}

\newif\ifusehyperref
\usehyperreftrue 
\ifusehyperref
 \usepackage[unicode,breaklinks]{hyperref}
 \usepackage[all]{hypcap}
 \hypersetup{backref,
             bookmarks,
             dvips,
             pdftitle=The limited role of mergers,
             pdfsubject=Astrophysics \& Cosmology,
             pdfkeywords={black hole physics -- galaxies: bulges},
             pdfdisplaydoctitle,
             colorlinks=true,
             citecolor=Blue,
             pdfauthor=Yohai Meiron,
             unicode}
\else
 \newcommand{\href}[2]{#2}
\fi

\def\kms{km\,s${}^{-1}$}

\title[The limited role of mergers]{The limited role of mergers in the black hole to bulge mass relation}

\author[C. G. Lahav, Y. Meiron and N. Soker]{Carmit G. Lahav,\thanks{E-mail: \href{mailto:carmitgo@tx.technion.ac.il}{carmitgo@tx.technion.ac.il} (CL); \newline\href{mailto:ym@physics.technion.ac.il}{ym@physics.technion.ac.il} (YM);\newline\href{mailto:soker@physics.technion.ac.il}{soker@physics.technion.ac.il} (NS);}
Yohai Meiron\footnotemark[1] and Noam Soker\footnotemark[1]\\
Department of Physics, Technion -- Israel Institute of Technology, Haifa 32000, Israel}

\begin{document}

\date{In original form 2012 April 5}

\pagerange{\pageref{firstpage}--\pageref{lastpage}} \pubyear{2012}

\maketitle

\label{firstpage}

\begin{abstract}
We examine the intrinsic scatter in the correlation between black hole masses and their host bulge masses, and find that it cannot be accounted for by mergers alone. A simple merger scenario of small galaxies leads to a proportionality relation between the late-time black hole and bulge masses, with intrinsic scatter (in linear scale) increasing along the ridge line of the relation as the square root of the mass. By examining a sample of 86 galaxies with well measured black hole masses, we find that the intrinsic scatter increases with mass more rapidly than expected from the merger-only scenario. We discuss the possibility that the feedback mechanism that operated during galaxy formation involved the presence of a cooling flow.
\end{abstract}


\purplethingy{Submitted}

\section{Introduction}
\label{sec:intro}

Relations between a supermassive black hole (SMBH) mass, $M_{\rm BH}$, and
properties of its host galaxy have been studied intensively for more than a
decade. Two of the galactic properties most commonly correlated with $M_{\rm BH}$
are the stellar mass of the spheroidal component (which we refer to as the
bulge), $M_{\rm G}$ (e.g. \citealt{KormendyRichstone1995, Magorrian_etal1998,
Laor2001, Hu2009, Graham2009}), and its stellar velocity dispersion, $\sigma$
(e.g. \citealt{Gebhardt_etal2000, MerrittFerrarese2001, Graham2008a,
Graham2008b, Hu2008, Shen_etal2008, Gultekin_etal2009}). These relations are
often assumed to have the form of a power law (i.e. linear when plotted on a
log-log scale).

However, due to the scarcity of data, large measurement error and intrinsic scatter, there is still no consensus on which galactic parameter has the best correlation with $M_{\rm BH}$. Some studies examine different
combinations of $\sigma$ and $M_{\rm G}$; \citet{Feoli_etal2010}, for example,
argued that $M_{\rm BH}$ is better correlated with the energy parameter,
$M_{\rm G} \sigma^2$, than with $M_{\rm G}$ or with $\sigma$ alone.
\citet{Soker2011} found that a momentum-like parameter $\mu \equiv M_{\rm G}
\sigma /c$ is well correlated with $M_{\rm BH}$, as predicted by the
penetrating-jet feedback model (\citealt{Soker2009}; although this mechanism is based on energy
balance).

Some studies consider the $M_{\rm BH}$--$M_{\rm G}$ relationship to be the
fundamental one, and further argue that mergers between many low mass galaxies
from an initially uncorrelated sample can lead to the observed correlation
\citep{Peng2007, Jahnke2011, Gaskell2011}. \citet{Peng2007} claimed
that the scatter decreases in logarithmic scale toward higher masses; motivated
by his view that `fine tuning' is required in feedback, he claimed that AGN feedback is neither necessary nor desirable to produce the observed correlation.
\citet{Jahnke2011} added star formation to the growth of galaxies by mergers,
and found the decrease in the relative scatter (i.e. in $M_{\rm BH}/M_{\rm G}$)
to be slower than when only mergers are considered; however, the decrease in the
scatter they obtained was still faster than the observed behaviour.

Following these arguments, we examine the scatter increase based on
mergers. We expect, as will be further explained in Section \ref{sec:scater},
that if only mergers cause BH growth, the intrinsic scatter of BH masses will increase as the
square root of number of mergers $n^{1/2}$ while the mass increases as $n$ (thus, the relative scatter will decrease as $n^{-1/2}$). Note that this refers to the intrinsic scatter on {\it linear scale} rather than logarithmic scale; the reason for discussing the intrinsic scatter in mass-mass relationships in linear scale will also be made clear in Section \ref{sec:scater}. We also note that in this work, $n$ of a particular galaxy is the number of building blocks that merged over time to form it (cf. `generation number', which is $\log_2 n$).

In the current paper, we examine the behaviour of the scatter in the $M_{\rm
BH}$--$M_{\rm G}$ relationship with mass in order to asses the importance of
mergers. The sample of galaxies is discussed in Section \ref{sec:sample}, and
the results in Section \ref{sec:scater}. In Section \ref{sec:summary} we discuss
the implications of the results to the feedback mechanism and give a brief
summary.

\begin{figure*}
{\includegraphics{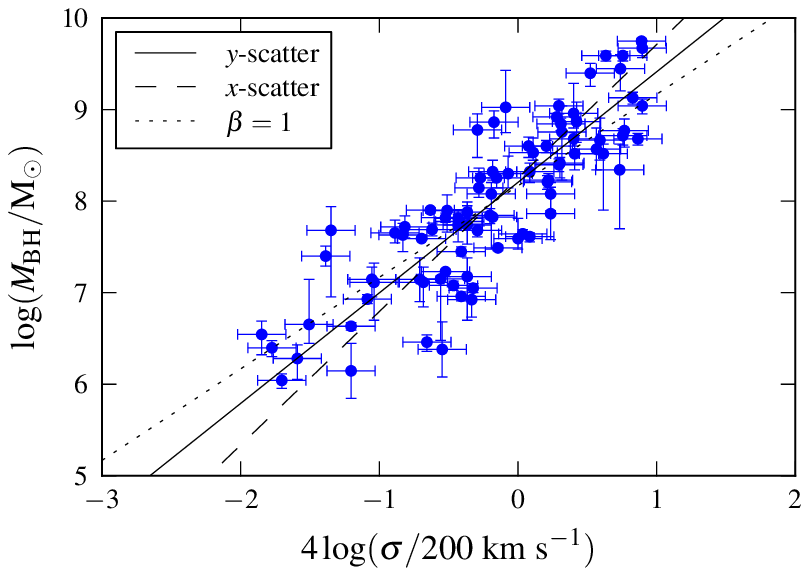}}
{\includegraphics{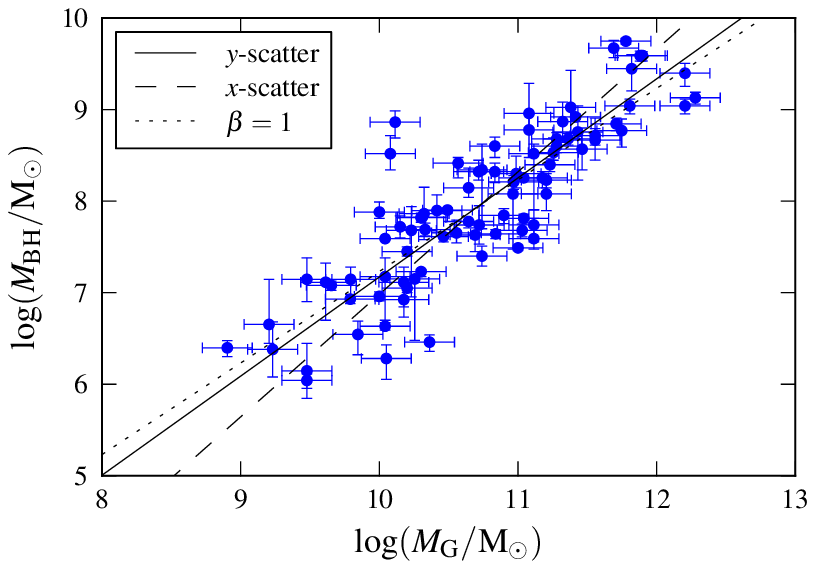}}
\caption{The $M_{\rm BH}$--$\sigma^4$ (left) and $M_{\rm BH}$--$M_{\rm G}$
(right) relations for our sample of 86 galaxies. Here $M_{\rm BH}$, $M_{\rm G}$, and $\sigma$ are the black hole mass, bulge mass, and stellar velocity dispersion of the galaxies, respectively, and the logarithm is base 10; the data are from Table \ref{tab:Galaxytabel}.
The
solid and dashed lines are the best-fitting power law relations using the $x$-
and $y$-scatter models respectively (see text). The dotted line in each panel
has a slope $\beta=1$.}
\label{fig:corsigmamg}
\end{figure*}

\section{The Sample}
\label{sec:sample}

Our sample of 86 galaxies was compiled from various sources which provided black
hole masses $M_{\rm BH}$, velocity dispersions $\sigma$, and bulge masses
$M_{\rm G}$. Most objects (60 galaxies) were taken from \cite{Graham2008b} with updated
values from \cite{Graham2011}. Values of $M_{\rm G}$ were taken from 
\citet{Feoli_etal2010} and references therein (table 1 there). Six galaxies with $M_{\rm BH}$ from \cite{Gultekin_etal2009} and $M_{\rm G}$ from table 2 of \citeauthor{Feoli_etal2010}; seven galaxies from \citeauthor{Hu2009} (\citeyear{Hu2009}, also listed in table 3 of \citeauthor{Feoli_etal2010}); five galaxies were taken from \cite{Greene_etal2010}; five and three SMBH masses were taken from \citet{Peterson_etal2004} and \citet{Bentz_etal2009} respectively, with $M_{\rm G}$ and $\sigma$ values from \citet{Wandel2002}. We adopted the value of 0.18 dex for the error in $M_{\rm G}$ as
\citeauthor{Feoli_etal2010}. The error in $\sigma$ was taken to be 10
per cent as in \cite{Graham2011}. Measurement errors in $M_{\rm BH}$ are as they appeared in the
sources listed above. We list all the data used in Table \ref{tab:Galaxytabel}.

In Fig. \ref{fig:corsigmamg} we present two correlations that have been
thoroughly studied for the past decade: $M_{\rm BH}$--$\sigma$ and $M_{\rm
BH}$--$M_{\rm G}$. Previous studies have firmly established that black hole mass
is tightly correlated with bulge mass and with the stellar velocity dispersion,
with power law relationships commonly assumed (e.g.
\citealt{KormendyRichstone1995, Magorrian_etal1998, Laor2001, Wandel2002,
Hu2009, Graham2009, Graham2011, Greene_etal2010, Gultekin_etal2009,
Gebhardt_etal2000, MerrittFerrarese2001, Tremaine_etal2002}).

\begin{figure}
\begin{center}
\includegraphics{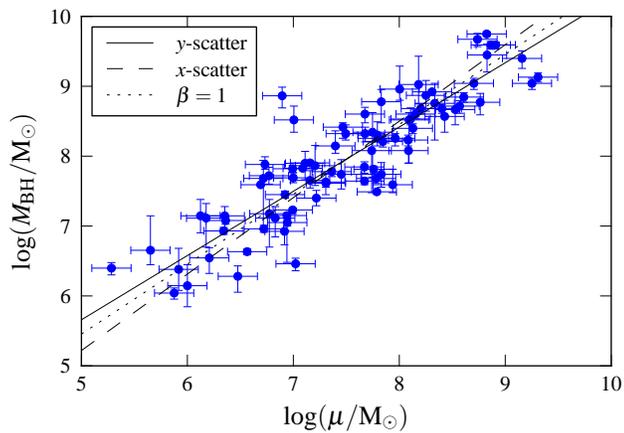}
\end{center}
\caption{The same as Fig. \ref{fig:corsigmamg} but for the $M_{\rm BH}$--$\mu$ relation, where $\mu \equiv M_{\rm G}\sigma/c$ (measured in solar masses). The
solid and dashed lines are the best-fitting power law relations using the $x$-
and $y$-scatter models respectively (see text). The dotted line in each panel
has a slope $\beta=1$.}
\label{fig:cor1}
\end{figure}

In addition to these widely examined relationships, other correlations have been considered in the
literature (e.g. \citealt{Mancini&Feoli2011}). \cite{Feoli_etal2010}, for
example, suggested that $M_{\rm BH}\sigma^2$ correlates better with $M_{\rm BH}$
than $M_{\rm G}$ or $\sigma$. \cite{Soker2011} presented the momentum parameter
$\mu\equiv M_{\rm G}\sigma/c$ as a galactic property which has a proportionality
relation with $M_{\rm BH}$ in the penetrating jet feedback mechanism; they
verified this relation using a sample of 49 galaxies. We repeat the analysis of
the latter ($M_{\rm BH}$--$\mu$) and show that this relation holds also for our
wider sample of 86 galaxies, as seen in Fig. \ref{fig:cor1}.

The best-fitting parameters for each correlation are calculated in three ways, or {\it scatter models}. A constant residual, $\varepsilon_0$, (the {\it intrinsic scatter}) is added (as an error) in one of three directions until the sum of square residuals is equal the number of degrees of freedom (the methodology is described in more detail in \citealt{Soker2011}). This method of estimating the intrinsic scatter is due to \citet{Tremaine_etal2002}, who considered scatter in just the SMBH mass. The three scatter models considered here give somewhat different results, but none is preferable as the true direction of the scatter depends on the physical
mechanism that leads to the correlation \citep{Novak}, and cannot be determined from the data. The three scatter models are: (1) $y$-scatter where the $\varepsilon_0$ is the residual
variance in $\log M_{\rm BH}$ (as in \citealt{Tremaine_etal2002} and most
studies thereafter); (2) $x$-scatter where $\varepsilon_0$ is the scatter in the
galactic property; and (3) {\it orthogonal scatter} where the scatter is added
orthogonally to the ridge line. In Table \ref{tab:slopes} we give three parameters for each of the three correlations in three scatter models. They are similar to those obtained in recent years (e.g. \citealt{Soker2011}, \citealt{Graham2011}).

\begin{table}
\caption{The best-fitting values and their uncertainties for the power law
relation $\log(M_{\rm BH})=\alpha + \beta\log P_{\rm G}$, where $P_{\rm G}$ is
the galactic parameter in question. The quantities $M_{\rm BH}$, $M_{\rm G}$ and
$\mu$ are in solar masses while $\sigma$ is in units of 200 \kms. The table
gives the results for the three models of intrinsic scatter discussed in the
text (further details regarding this calculation can be found in
\citealt{Soker2011}).\label{tab:slopes}}
\centering
\begin{tabular}{clrrr}
\hline\hline
Correlation & Model & $\alpha\pm\delta\alpha$ & $\beta\pm\delta\beta$ & $\varepsilon_0$\\\hline
\multirow{3}{*}{$M_{\rm BH}$--$\mu$}       & $x$-scatter        & $-0.25\pm0.47$ & $1.09\pm0.06$ & 0.35\\
                                           & $y$-scatter        & $ 1.06\pm0.40$ & $0.92\pm0.05$ & 0.35\\
                                           & {\it orth}-scatter & $ 0.42\pm0.47$ & $1.00\pm0.06$ & 0.35\\\hline
\multirow{3}{*}{$M_{\rm BH}$--$M_{\rm G}$} & $x$-scatter        & $-6.40\pm0.91$ & $1.34\pm0.08$ & 0.32\\
                                           & $y$-scatter        & $-3.65\pm0.74$ & $1.08\pm0.07$ & 0.38\\
                                           & {\it orth}-scatter & $-5.93\pm0.97$ & $1.29\pm0.09$ & 0.34\\\hline
\multirow{3}{*}{$M_{\rm BH}$--$\sigma^4$}  & $x$-scatter        & $ 8.25\pm0.06$ & $1.46\pm0.09$ & 0.28\\
                                           & $y$-scatter        & $ 8.21\pm0.05$ & $1.21\pm0.08$ & 0.37\\
                                           & {\it orth}-scatter & $ 8.25\pm0.04$ & $1.44\pm0.10$ & 0.29\\\hline
\end{tabular}
\end{table}

\section{The Scatter}
\label{sec:scater}

We first clarify two confusing issues. One is that the (total) `scatter' of the data points (quantified by the root mean square of the residuals from the best fitting curve) is not the same as the intrinsic scatter, which is a measure of the natural spread of the data. The total scatter has an intrinsic component but is larger due to measurement errors. From here on, we will denote the rms of the residuals by $\sigma$ (not to be confused with the stellar velocity dispersion, which will not be mentioned further) and the intrinsic scatter by $\varepsilon_0$ as in Section \ref{sec:sample}. The second issue is that while in other works (as well as the previous Section of this paper) the intrinsic scatter was estimated from the logarithmic data and was thus dimensionless, from here on, we will consider the scatter (extrinsic and intrinsic) in linear scale.

We now examine the hypothesis that the correlations are mainly due to mergers.
Let us consider a simple
scenario of growth through mergers: the building blocks are galaxies with equal
initial bulge mass $M_{{\rm G},0}$, and initial black hole masses with some
distribution with an expected value $M_{{\rm BH},0}$ and variance $s_0^2$. After
$n \gg 1$ mergers, the bulge mass is $M_{\rm G}=n M_{{\rm G},0}$, and the black
hole masses are normally distributed around $M_{\rm BH}=n M_{{\rm BH},0}$ with
variance $s^2 = n s_0^2$. The central limit theorem asserts that $M_{\rm BH}$
will be distributed normally for a given $M_{\rm G}$ (or equivalently, a given
$n$), independently of the initial distribution, as long as $n \gg 1$.
Thus, this model leads to a proportionality relation between black hole and
bulge mass: $M_{\rm BH} = M_{\rm G} (M_{{\rm BH},0} / M_{{\rm G},0})$, with
intrinsic scatter which increases as $\sqrt{M_{\rm G}}$.
This result, while derived from a very simplistic model, is consistent with the results of \citet{Hirschmann}, who studied the evolution of the intrinsic scatter with redshift using cosmological halo merger trees \citep{Genel}; this is because the only `physics' involved is convergence toward a Gaussian distribution.

Since $M_{\rm BH}
\propto M_{\rm G}$, then obviously the square roots of the masses also satisfy a
proportionality relation: $\sqrt{M_{\rm BH}} \propto \sqrt{M_{\rm G}}$. The
scatter $\epsilon_0$ of $\sqrt{M_{\rm BH}}$ for a given $\sqrt{M_{\rm G}}$ can be
calculated by error propagation:
\begin{equation}
\epsilon_0 =
\left( \frac {{\rm d}\sqrt{M_{\rm BH}}}{{\rm d} M_{\rm BH}} \right) s
= \frac{s}{2 \sqrt{M_{\rm BH}}}
= \frac{\sqrt{n} s_0}{2 \sqrt{n M_{{\rm BH},0}}}
= \frac{s_0}{2 \sqrt{M_{{\rm BH},0}}}
= {\rm const.}
\label{eq:Delta1}
\end{equation}
Thus, this model for galaxy--SMBH co-evolution, which is based on mergers alone
predicts an $M_{\rm BH}^{1/2}$--$M_{\rm G}^{1/2}$ relation with scatter that
does not depend on mass. Since the black hole masses (or their square root) form
a normal distributed for a given bulge mass, this is a `$y$'-type scatter;
relaxing the assumption that the initial bulge masses are identical would lead
to a scatter in a different direction.

\begin{figure*}
\includegraphics{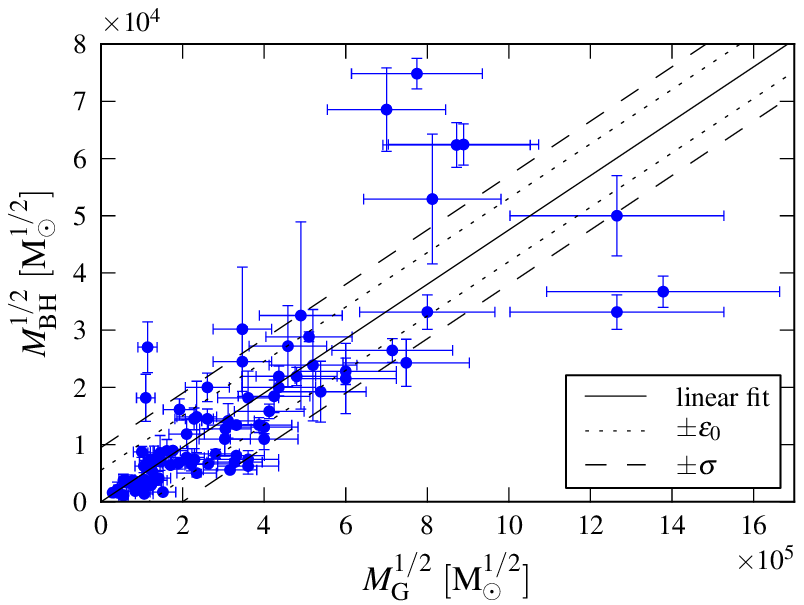}
\includegraphics{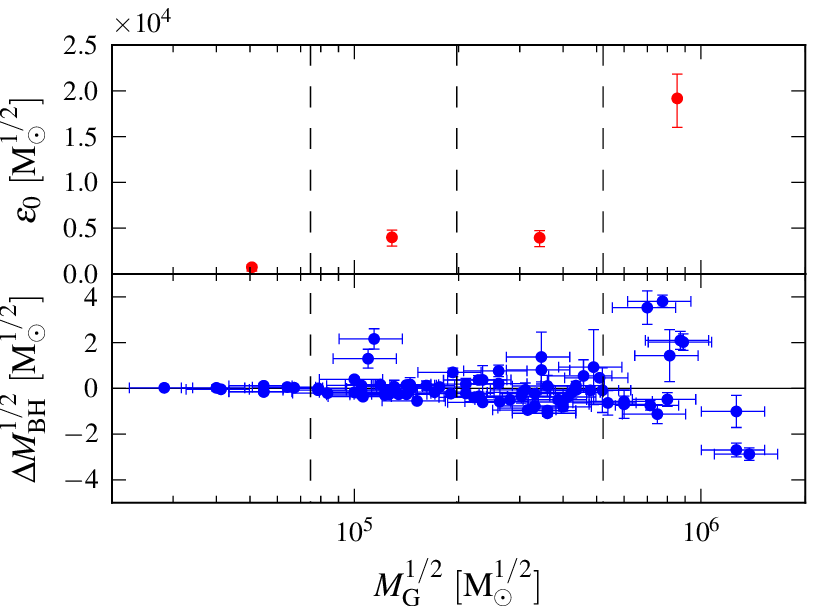}
\caption{Left: The correlation of $M_{\rm BH}^{1/2}$ and $M_{\rm G}^{1/2}$,
where $M_{\rm BH}$ and  $M_{\rm G}$ are the SMBH and host galaxy bulge masses,
respectively. The solid line represents the linear fit, the dashed lines delimit a region of one standard deviation (residual rms) from the ridge line, and the dotted lines represent the intrinsic scatter of all data points (see text). Right: the bottom panel shows the vertical distance of each data point from the ridge line versus $M_{\rm G}^{1/2}$; the upper panel shows the intrinsic scatter calculated for each of the four bins.}
\label{fig:cor2}
\end{figure*}

The hypothesis that the scatter of $M_{\rm BH}^{1/2}$ is constant can be easily
tested even with current measurements. Fig. \ref{fig:cor2} (left) shows the
$M_{\rm BH}^{1/2}$--$M_{\rm G}^{1/2}$ correlation (the data is the same as Fig. \ref{fig:corsigmamg}, but on linear rather than logarithmic scale). The solid line is the best fitting proportionality relation (zero intercept); the dotted and dashed lines are the intrinsic and total scatter on $M_{\rm BH}$ for all the data, respectively.

The residuals are plotted in the bottom right panel of Fig. \ref{fig:cor2}. It is easy to see that the total scatter around the ridge line increases with bulge mass, but the intrinsic scatter needs be calculated.
We divided the dataset into four bins in $M_{\rm G}$ with equal logarithmic width, containing (from low to high mass) 8, 29, 35 and 14 objects. For each bin we calculated the intrinsic scatter on $M_{\rm BH}$ required to bring the reduced sum of square residuals (from the line calculated from the entire dataset) to 1. The errors on $\epsilon_0$ are calculated from the shape of the $\chi^2$ distribution as explained in \citet{Soker2011}.
As the upper right panel of Fig. \ref{fig:cor2} shows, the intrinsic scatter increases toward higher masses, in contradiction to the prediction of the mergers-only model discussed above.

\section{Discussion and Summary}
\label{sec:summary}

Our objective was to examine the claim that the SMBH mass to bulge mass ($M_{\rm
BH}$--$M_{\rm G}$) correlation is predominantly a result of mergers of low mass
galaxies.
The merging process leads to an intrinsic scatter in the
$M_{\rm BH}^{1/2}$--$M_{\rm G}^{1/2}$ relation which is uniform as given by equation
(\ref{eq:Delta1}). We checked this prediction with a sample of 86 galaxies,
listed in Table \ref{tab:Galaxytabel}. Some properties of the sample are
presented in Figs. \ref{fig:corsigmamg} and \ref{fig:cor1}, and in Table
\ref{tab:slopes}. In Fig. \ref{fig:cor2} we show the
$M_{\rm BH}^{1/2}$--$M_{\rm G}^{1/2}$ correlation and the residuals.

Our main result is that the intrinsic scatter increases with mass more rapidly than
expected in a merger-only scenario. Since mergers occur between all types of
galaxies, we include all types of galaxies in our study. However, we checked our
analysis for ellipticals, spirals, barred, unbarred, AGNs and inactive galaxies
and found the main conclusion to hold for each subgroup separately.

While biasing is always a worry when considering these correlations (especially given the scarcity of data), it
does not seem like a selection effect can artificially produce the
result of our Fig. \ref{fig:cor2}: the pure merger scenario is only true when the
residuals are scattered uniformly; this uniform scatter cannot be
smaller than what we see at the high mass end, which is $2\times 10^4$ in units
of ${\rm M}_\odot^{1/2}$. Missing galaxies at the high end (e.g.
small $M_{\rm BH}$) can only increase the scatter.
At the low mass end, it is only reasonable that we `miss' black holes
with masses below the ridge line, but even if they exist, they cannot be
scattered over $2\times 10^4~{\rm M}_\odot^{1/2}$ since at the low mass end, $M_{\rm BH}$
is smaller than that. Also, note the recent work by \citet{Gultekin_etal2011}, who investigated the possibility that selection
effects (namely missing very low mass SMBHs) bias the measurements, and
reached the conclusion that it is not likely that the published
relations are biased. Our sample was not selected in any special way
and largely overlaps with theirs.

Another argument why the pure merger scenario is discrepant with the observations is that the $M_{\rm
BH}$--$M_{\rm G}$ relationship might not be a true proportionality relation, but have higher order terms. This is indicated by the fact that in two out of the three fitting methods (scatter models) for this relation in logarithmic scale, a slope of 1 (indicating a proportionality relation rather than an arbitrary power law) was more than three standard deviations from the best fitting slope (see Table \ref{tab:slopes}).

We do not dispute the claim that mergers influence the $M_{\rm BH}$--$P_{\rm G}$
relations, where $P_{\rm G}$ represents any property of the host galaxy. We only
argue that based on our results, mergers cannot be the dominant cause of
correlation. Rather, we expect this correlation to be determined mainly by a
feedback mechanism that operates on all scales, from small galaxies to galaxy
clusters.

Let us demonstrate a simple plausible implication of the hypothesis that the
same feedback mechanism operates on all scales. It is hard to observe the
processes that took place during galaxy formation at the high redshift universe,
and many models exist, based on radiation and/or jets from AGNs and/or star
formation, e.g. \cite{SilkRees1998}, \cite{King2003}, \cite{Soker2009}, and
\cite{SilkNusser2010}. The situation is much better with cooling flow clusters,
that show both star formation and AGN activity. The central galaxy of the Virgo
cluster, M87 (NGC 4486), and the central galaxy of Fornax, NGC 1399, 
are two cooling flow clusters in our sample.

We take the energy in jets from AGN outbursts to be due to accretion on to the
SMBH with an efficiency $\eta_j$: $E_{\rm jets} = \eta_j M_{\rm acc} c^2$. The
jets energy is inferred from the energy required to inflate X-ray deficient
cavities (bubbles). We here assume that the jet's energy source is the accreted
mass (but see \citealt{McNamara2011}). Under this assumption,
\cite{McNamara2011} find the median value of the molecular mass in the central
galaxy to mass accreted during AGN outburst ratio to be $M_{\rm mol}/ M_{\rm
acc} \simeq 700  (\eta_j /0.04)^{-1}$, where the scaling of $\eta_j=0.04$ is
based on the results of \cite{Soker2011}. This is very close to the ratio
$M_{\rm G}/M_{\rm BH} = 590 \pm 70$ found in our sample. The feedback process in
cooling flow clusters seems to be able to produce the $M_{\rm BH}$--$M_{\rm G}$
correlation. This might hint that the feedback mechanism that operated during
galaxy formation involved the presence of a cooling flow, as suggested in an
earlier paper \citep{Soker2010}.

\medskip

This research was supported by the Asher Fund for Space Research at the Technion, and the Israel Science Foundation.

\newcommand{\mnras}{MNRAS}
\newcommand{\apj}{ApJ}
\newcommand{\apjl}{ApJ}

\onecolumn

\begin{deluxetable}{lrrrrr}
\tablecolumns{6}
\small
\tablewidth{0pt}
\tablecaption{Our sample of 86 galaxies\label{tab:Galaxytabel}}
\tablehead{
\colhead{Galaxy} &
\colhead{$\sigma$} &
\colhead{$M_{\rm BH}$-high} &
\colhead{$M_{\rm BH}$-low} &
\colhead{$M_{\rm BH}$} &
\colhead{$M_{\rm G}$}
\\
\colhead{} &
\colhead{[\kms]} &
\colhead{[$M_\odot$]} &
\colhead{[$M_\odot$]} &
\colhead{[$M_\odot$]} &
\colhead{[$M_\odot$]}
}
\startdata
Milky Way&$100$&$4.7\times10^{6}$&$3.9\times10^{6}$&$4.3\times10^{6}$&$1.1\times10^{10}$\\
NGC 821&$200$&$6.5\times10^{7}$&$3.0\times10^{7}$&$3.9\times10^{7}$&$1.3\times10^{11}$\\
NGC 2778&$162$&$2.4\times10^{7}$&$5.0\times10^{6}$&$1.5\times10^{7}$&$1.1\times10^{10}$\\
NGC 3379&$209$&$5.0\times10^{8}$&$3.0\times10^{8}$&$4.0\times10^{8}$&$6.8\times10^{10}$\\
NGC 3384&$148$&$1.8\times10^{7}$&$1.5\times10^{7}$&$1.7\times10^{7}$&$2.0\times10^{10}$\\
NGC 3608&$192$&$3.1\times10^{8}$&$1.4\times10^{8}$&$2.0\times10^{8}$&$9.7\times10^{10}$\\
NGC 4291&$285$&$4.2\times10^{8}$&$8.0\times10^{7}$&$3.3\times10^{8}$&$1.3\times10^{11}$\\
NGC 4473&$179$&$1.6\times10^{8}$&$3.0\times10^{7}$&$1.2\times10^{8}$&$9.2\times10^{10}$\\
NGC 4486&$334$&$6.0\times10^{9}$&$5.2\times10^{9}$&$5.6\times10^{9}$&$6.0\times10^{11}$\\
NGC 4564&$157$&$6.3\times10^{7}$&$5.1\times10^{7}$&$6.0\times10^{7}$&$4.4\times10^{10}$\\
NGC 4649&$335$&$5.7\times10^{9}$&$3.7\times10^{9}$&$4.7\times10^{9}$&$4.9\times10^{11}$\\
NGC 4697&$171$&$2.0\times10^{8}$&$1.7\times10^{8}$&$1.8\times10^{8}$&$1.1\times10^{11}$\\
NGC 5128&$120$&$6.2\times10^{7}$&$3.5\times10^{7}$&$4.5\times10^{7}$&$3.6\times10^{10}$\\
NGC 5845&$238$&$3.0\times10^{8}$&$1.8\times10^{8}$&$2.6\times10^{8}$&$3.7\times10^{10}$\\
Circinus&$75$&$1.3\times10^{6}$&$9.0\times10^{5}$&$1.1\times10^{6}$&$3.0\times10^{9}$\\
Cygnus A&$270$&$3.2\times10^{9}$&$1.8\times10^{9}$&$2.5\times10^{9}$&$1.6\times10^{12}$\\
NGC 221&$72$&$3.0\times10^{6}$&$2.0\times10^{6}$&$2.5\times10^{6}$&$8.0\times10^{8}$\\
NGC 224&$170$&$2.3\times10^{8}$&$1.1\times10^{8}$&$1.4\times10^{8}$&$4.4\times10^{10}$\\
NGC 1023&$204$&$4.9\times10^{7}$&$3.9\times10^{7}$&$4.4\times10^{7}$&$6.9\times10^{10}$\\
NGC 1300&$229$&$1.4\times10^{8}$&$3.8\times10^{7}$&$7.3\times10^{7}$&$2.1\times10^{10}$\\
NGC 1399&$329$&$5.5\times10^{8}$&$4.1\times10^{8}$&$4.8\times10^{8}$&$2.3\times10^{11}$\\
NGC 2787&$210$&$4.5\times10^{7}$&$3.6\times10^{7}$&$4.1\times10^{7}$&$2.9\times10^{10}$\\
NGC 3031&$162$&$9.8\times10^{7}$&$6.5\times10^{7}$&$7.6\times10^{7}$&$1.0\times10^{10}$\\
NGC 3079&$146$&$4.8\times10^{6}$&$1.2\times10^{6}$&$2.4\times10^{6}$&$1.7\times10^{9}$\\
NGC 3115&$252$&$1.9\times10^{9}$&$6.3\times10^{8}$&$9.1\times10^{8}$&$1.2\times10^{11}$\\
NGC 3227&$133$&$2.4\times10^{7}$&$8.0\times10^{6}$&$1.4\times10^{7}$&$3.0\times10^{9}$\\
NGC 3245&$210$&$2.6\times10^{8}$&$1.6\times10^{8}$&$2.1\times10^{8}$&$6.8\times10^{10}$\\
NGC 3377&$139$&$8.5\times10^{7}$&$7.4\times10^{7}$&$8.0\times10^{7}$&$3.1\times10^{10}$\\
NGC 3998&$305$&$4.2\times10^{8}$&$5.0\times10^{7}$&$2.2\times10^{8}$&$5.5\times10^{10}$\\
NGC 4151&$156$&$7.2\times10^{7}$&$5.8\times10^{7}$&$6.5\times10^{7}$&$1.1\times10^{11}$\\
NGC 4258&$134$&$4.0\times10^{7}$&$3.8\times10^{7}$&$3.9\times10^{7}$&$1.1\times10^{10}$\\
NGC 4261&$309$&$6.2\times10^{8}$&$4.1\times10^{8}$&$5.2\times10^{8}$&$3.6\times10^{11}$\\
NGC 4342&$253$&$5.2\times10^{8}$&$2.2\times10^{8}$&$3.3\times10^{8}$&$1.2\times10^{10}$\\
NGC 4374&$281$&$8.1\times10^{8}$&$2.8\times10^{8}$&$4.6\times10^{8}$&$3.6\times10^{11}$\\
NGC 4459&$178$&$8.3\times10^{7}$&$5.7\times10^{7}$&$7.0\times10^{7}$&$7.9\times10^{10}$\\
NGC 4486a&$110$&$2.1\times10^{7}$&$5.0\times10^{6}$&$1.3\times10^{7}$&$4.1\times10^{9}$\\
NGC 4596&$149$&$1.2\times10^{8}$&$4.6\times10^{7}$&$7.9\times10^{7}$&$2.6\times10^{10}$\\
NGC 4945&$100$&$2.8\times10^{6}$&$7.0\times10^{5}$&$1.4\times10^{6}$&$3.0\times10^{9}$\\
NGC 5077&$255$&$1.2\times10^{9}$&$4.4\times10^{8}$&$7.4\times10^{8}$&$2.1\times10^{11}$\\
NGC 5252&$190$&$2.7\times10^{9}$&$5.6\times10^{8}$&$1.1\times10^{9}$&$2.4\times10^{11}$\\
NGC 6251&$311$&$7.9\times10^{8}$&$3.9\times10^{8}$&$5.9\times10^{8}$&$5.6\times10^{11}$\\
NGC 7052&$277$&$6.3\times10^{8}$&$2.2\times10^{8}$&$3.7\times10^{8}$&$2.9\times10^{11}$\\
NGC 7582&$156$&$8.1\times10^{7}$&$3.6\times10^{7}$&$5.5\times10^{7}$&$1.3\times10^{11}$\\
NGC 2974&$227$&$2.0\times10^{8}$&$1.4\times10^{8}$&$1.7\times10^{8}$&$1.6\times10^{11}$\\
NGC 3414&$237$&$2.9\times10^{8}$&$2.1\times10^{8}$&$2.5\times10^{8}$&$1.7\times10^{11}$\\
NGC 4552&$252$&$5.6\times10^{8}$&$4.0\times10^{8}$&$4.8\times10^{8}$&$1.9\times10^{11}$\\
NGC 4621&$225$&$4.6\times10^{8}$&$3.4\times10^{8}$&$4.0\times10^{8}$&$1.9\times10^{11}$\\
NGC 5813&$239$&$8.1\times10^{8}$&$5.9\times10^{8}$&$7.0\times10^{8}$&$5.1\times10^{11}$\\
NGC 5846&$237$&$1.3\times10^{9}$&$9.0\times10^{8}$&$1.1\times10^{9}$&$6.4\times10^{11}$\\
Abell 1836&$309$&$4.3\times10^{9}$&$3.4\times10^{9}$&$3.9\times10^{9}$&$7.9\times10^{11}$\\
Abell 3565&$335$&$1.3\times10^{9}$&$9.0\times10^{8}$&$1.1\times10^{9}$&$1.6\times10^{12}$\\
NGC 1068&$165$&$8.7\times10^{6}$&$5.4\times10^{6}$&$8.4\times10^{6}$&$1.5\times10^{10}$\\
IC 1459&$306$&$4.0\times10^{9}$&$1.6\times10^{9}$&$2.8\times10^{9}$&$6.6\times10^{11}$\\
NGC 2748&$92$&$8.7\times10^{7}$&$9.0\times10^{6}$&$4.8\times10^{7}$&$1.7\times10^{10}$\\
NGC 4350&$181$&$9.7\times10^{8}$&$4.9\times10^{8}$&$7.3\times10^{8}$&$1.3\times10^{10}$\\
NGC 4486B&$169$&$9.0\times10^{8}$&$3.0\times10^{8}$&$6.0\times10^{8}$&$1.2\times10^{11}$\\
NGC 4742&$109$&$1.9\times10^{7}$&$9.0\times10^{6}$&$1.4\times10^{7}$&$6.2\times10^{9}$\\
NGC 7332&$135$&$1.9\times10^{7}$&$7.0\times10^{6}$&$1.3\times10^{7}$&$1.5\times10^{10}$\\
NGC 7457&$69$&$4.9\times10^{6}$&$2.1\times10^{6}$&$3.5\times10^{6}$&$7.0\times10^{9}$\\
NGC 7469&$153$&$1.3\times10^{7}$&$1.1\times10^{7}$&$1.2\times10^{7}$&$4.5\times10^{9}$\\
NGC 1194&$148$&$7.4\times10^{7}$&$5.9\times10^{7}$&$6.6\times10^{7}$&$2.0\times10^{10}$\\
NGC 2960&$166$&$1.3\times10^{7}$&$1.0\times10^{7}$&$1.1\times10^{7}$&$1.6\times10^{10}$\\
NGC 4388&$107$&$9.5\times10^{6}$&$7.6\times10^{6}$&$8.5\times10^{6}$&$6.2\times10^{9}$\\
NGC 6264&$158$&$3.2\times10^{7}$&$2.5\times10^{7}$&$2.8\times10^{7}$&$1.6\times10^{10}$\\
NGC 6323&$158$&$1.0\times10^{7}$&$8.1\times10^{6}$&$9.1\times10^{6}$&$1.0\times10^{10}$\\
NGC 3585&$213$&$4.9\times10^{8}$&$2.8\times10^{8}$&$3.4\times10^{8}$&$1.8\times10^{11}$\\
NGC 3607&$229$&$1.6\times10^{8}$&$7.9\times10^{7}$&$1.2\times10^{8}$&$1.6\times10^{11}$\\
NGC 4026&$180$&$2.8\times10^{8}$&$1.7\times10^{8}$&$2.1\times10^{8}$&$5.2\times10^{10}$\\
NGC 4594&$240$&$1.1\times10^{9}$&$1.7\times10^{8}$&$5.7\times10^{8}$&$2.7\times10^{11}$\\
NGC 5576&$183$&$2.1\times10^{8}$&$1.4\times10^{8}$&$1.8\times10^{8}$&$1.5\times10^{11}$\\
NGC 4303&$84$&$1.4\times10^{7}$&$2.8\times10^{6}$&$4.5\times10^{6}$&$1.6\times10^{9}$\\
NGC 524&$235$&$8.9\times10^{8}$&$7.9\times10^{8}$&$8.3\times10^{8}$&$2.6\times10^{11}$\\
NGC 1316&$226$&$1.9\times10^{8}$&$1.3\times10^{8}$&$1.6\times10^{8}$&$9.3\times10^{10}$\\
NGC 2549&$145$&$1.5\times10^{7}$&$3.0\times10^{6}$&$1.4\times10^{7}$&$1.8\times10^{10}$\\
PGC 49940&$288$&$4.4\times10^{9}$&$3.4\times10^{9}$&$3.9\times10^{9}$&$7.6\times10^{11}$\\
IC 4296&$322$&$1.5\times10^{9}$&$1.1\times10^{9}$&$1.3\times10^{9}$&$1.9\times10^{12}$\\
NGC 3393&$184$&$3.3\times10^{7}$&$2.9\times10^{7}$&$3.1\times10^{7}$&$1.0\times10^{11}$\\
IC 2560&$137$&$3.5\times10^{6}$&$2.3\times10^{6}$&$2.9\times10^{6}$&$2.3\times10^{10}$\\
NGC 3516 &$124$&$5.7\times10^{7}$&$2.8\times10^{7}$&$4.3\times10^{7}$&$4.9\times10^{10}$\\
NGC 4051 &$80$&$2.7\times10^{6}$&$1.1\times10^{6}$&$1.9\times10^{6}$&$1.1\times10^{10}$\\
NGC 5548&$180$&$7.0\times10^{7}$&$6.5\times10^{7}$&$6.7\times10^{7}$&$2.0\times10^{10}$\\
3C 120&$162$&$8.9\times10^{7}$&$3.4\times10^{7}$&$5.5\times10^{7}$&$5.2\times10^{10}$\\
Mrk 79&$125$&$6.9\times10^{7}$&$4.0\times10^{7}$&$5.2\times10^{7}$&$1.4\times10^{10}$\\
Mrk 110 &$90$&$3.2\times10^{7}$&$1.9\times10^{7}$&$2.5\times10^{7}$&$5.5\times10^{10}$\\
Mrk 590&$169$&$5.6\times10^{7}$&$4.1\times10^{7}$&$4.8\times10^{7}$&$1.1\times10^{11}$\\
Mrk 817 &$140$&$5.8\times10^{7}$&$4.2\times10^{7}$&$4.9\times10^{7}$&$2.1\times10^{10}$\\
\enddata
\tablenotetext{{}}{
The fractional error in $\sigma$ is assumed to be 10 per cent as in
\cite{Graham2011}; the measurement error in $M_{\rm G}$ is 0.18 dex as in
\citet{Feoli_etal2010}. $M_{\rm BH}$-high and -low are the upper and lower limits of
the SMBH mass, respectively, according to the data available in the the references as detailed
in text. The first column on the right gives the bulge mass,
which is the stellar mass of the spheroidal component.}
\end{deluxetable}

\label{lastpage}


\begin{thebibliography}{}\addcontentsline{toc}{section}{References}


\bibitem[Bentz et al.(2009)]{Bentz_etal2009} Bentz, M.~C., Peterson, 
B.~M., Pogge, R.~W., \& Vestergaard, M.\ 2009, \apjl, 694, L166 

\bibitem[Feoli et al.(2011)]{Feoli_etal2010} Feoli, A., Mancini, L., 
Marulli, F., 
\& van den Bergh, S.\ 2011, General Relativity and Gravitation, 43, 1007 


\bibitem[Gaskell(2011)]{Gaskell2011} Gaskell, C.~M.\ 2011, arXiv:1111.2067

\bibitem[Gebhardt et al.(2000)]{Gebhardt_etal2000} Gebhardt, K., et al.\
2000, \apjl, 539, L13

\bibitem[Genel et al.(2009)]{Genel} Genel, S., Genzel, R., 
Bouch{\'e}, N., Naab, T., \& Sternberg, A.\ 2009, \apj, 701, 2002 

\bibitem[Graham (2008a)]{Graham2008a} Graham, A. W., 2008a, ApJ, 680, 143

\bibitem[Graham (2008b)]{Graham2008b} Graham, A. W., 2008b, PASA, 25, 167

\bibitem[Graham  \& Spitler (2009)]{Graham2009} Graham, A. W., \& Spitler, L. R.\
    2009, MNRAS, 397, 2148

\bibitem[Graham et al.(2011)]{Graham2011} Graham, A.~W., Onken, 
C.~A., Athanassoula, E., \& Combes, F.\ 2011, \mnras, 412, 2211

\bibitem[Greene et al.(2010)]{Greene_etal2010} Greene, J.~E., Peng, 
C.~Y., Kim, M., et al.\ 2010, \apj, 721, 26 

\bibitem[G{\"u}ltekin et al.(2009)]{Gultekin_etal2009} G{\"u}ltekin, K.,
et al.\ 2009, ApJ, 698, 198

\bibitem[G{\"u}ltekin et al.(2011)]{Gultekin_etal2011} G{\"u}ltekin, K., 
Tremaine, S., Loeb, A., \& Richstone, D.~O.\ 2011, \apj, 738, 17

\bibitem[Hirschmann et al.(2010)]{Hirschmann} Hirschmann, M., Khochfar, S., Burkert, A., et al.\ 2010, \mnras, 407, 1016 

\bibitem[Hu(2008)]{Hu2008} Hu, J.\  2008, MNRAS 386, 2242

\bibitem[Hu(2009)]{Hu2009} Hu, J.\ 2009, arXiv:0908.2028

\bibitem[Jahnke \& Macci{\`o}(2011)]{Jahnke2011} Jahnke, K., \& Macci{\`o}, A.~V.\ 2011, \apj, 734, 92:

\bibitem[King(2003)]{King2003} King, A.\ 2003, ApJl, 596, L27

\bibitem[Kormendy
\& Richstone(1995)]{KormendyRichstone1995} Kormendy, J., \& Richstone, D.\ 1995, ARAA, 33, 581

\bibitem[Laor(2001)]{Laor2001} Laor, A.\ 2001, ApJ, 553, 677

\bibitem[Magorrian et al.(1998)]{Magorrian_etal1998} Magorrian, J., et
al.\ 1998, AJ, 115, 2285

\bibitem[Mancini \& Feoli(2011)]{Mancini&Feoli2011} Mancini, L., \& Feoli, A.\ 2011, arXiv:1110.3542

\bibitem[McNamara et al.(2011)]{McNamara2011} McNamara, B.~R., Rohanizadegan, M., \& Nulsen, P.~E.~J.\ 2011, \apj, 727, 39

\bibitem[Merritt
\& Ferrarese(2001)]{MerrittFerrarese2001} Merritt, D., \& Ferrarese, L.\ 2001, ApJ, 547, 140

\bibitem[Novak et al.(2006)]{Novak} Novak, G.~S., Faber, 
S.~M., \& Dekel, A.\ 2006, \apj, 637, 96 

\bibitem[Peng(2007)]{Peng2007} Peng, C.~Y.\ 2007, \apj, 671, 1098

\bibitem[Peterson et al.(2004)]{Peterson_etal2004} Peterson, B.~M.,
Ferrarese, L., Gilbert, K.~M., et al.\ 2004, \apj, 613, 682

\bibitem[Shen et al.(2008)]{Shen_etal2008} Shen, J., Vanden Berk,
D.~E., Schneider, D.~P., \& Hall, P.~B.\ 2008, AJ, 135, 928

\bibitem[Silk \& Nusser(2010)]{SilkNusser2010} Silk, J., \& Nusser, A.\ 2010, arXiv:1004.0857

\bibitem[Silk \& Rees(1998)]{SilkRees1998} Silk, J., \& Rees, M.~J.\ 1998, AAP, 331, L1

\bibitem[Soker(2009)]{Soker2009} Soker, N.\ 2009, MNRAS, 398, L41

\bibitem[Soker(2010)]{Soker2010} Soker, N.\ 2010, \mnras, 407, 2355

\bibitem[Soker \& Meiron(2011)]{Soker2011} Soker, N., \& Meiron, Y.\ 2011, \mnras, 411, 1803

\bibitem[Tremaine et al.(2002)]{Tremaine_etal2002} Tremaine, S., et al.
\ 2002, ApJ, 574, 740

\bibitem[Wandel(2002)]{Wandel2002} Wandel, A.\ 2002, \apj, 565, 
762 

\end{thebibliography}
\end{document}